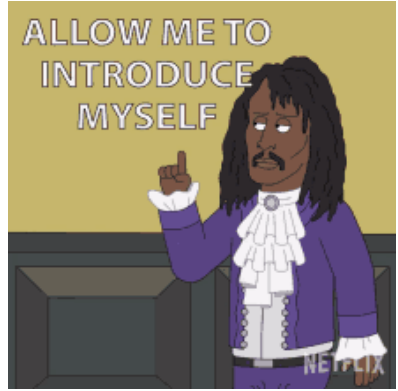

Dwayne Woods
Purdue University
Dwoods2@purdue.edu

# The Sponge Cake Dilemma over the Nile: Achieving Fairness in Resource Allocation with Cake Cutting Algorithms

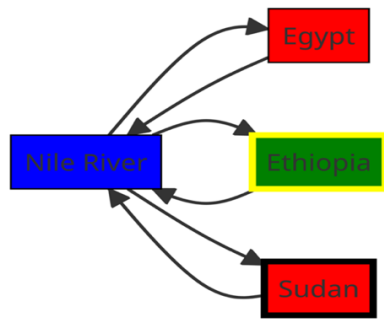

**Abstract**: This paper examines the complex dispute over water allocation in the Nile River basin through an integrated framework combining normative and positive perspectives. It applies political philosopher John Rawls' theory of justice to argue that all riparian states inherently deserve reasonable water access. Additionally, it utilizes mathematician Hugo Steinhaus's "cake-cutting" model to show that unchecked competition could enable monopoly. While Rawls defines ideally fair ends, cake-cutting provides means attuned to actual strategic behaviors. By incorporating AI and algorithmic solutions, particularly fair division algorithms, this paper explores how these technologies can offer practical pathways to achieving equitable and sustainable resource allocation. The strategic dimension of Ethiopia's Grand Renaissance Dam project is analyzed through game theory models, proposing an "I cut, you choose" approach to balance asymmetries. The paper concludes that integrating AI into conflict resolution frameworks can uphold equitable principles while recognizing that implementation pathways depend on strategic accommodation.

# I. Introduction:

The Nile River basin faces a long-standing dispute over water allocation among its riparian states, particularly Egypt, Sudan, and Ethiopia (Swain 2011). Despite numerous attempts at negotiation and mediation, the conflict remains unresolved, highlighting the need for innovative approaches to conflict resolution. This article examines the Nile water dispute through the lenses of game theory, fair division algorithms, and AI-driven solutions, aiming to provide new insights into the challenges of achieving a resolution and exploring potential pathways forward.

Traditional cooperative game theory assumes binding agreements and communication among players and has struggled to offer stable solutions to the Nile conflict. Even with Nash equilibrium, representing a state where no player can unilaterally improve their outcome, cooperative solutions remain elusive due to divergent interests and the lack of effective enforcement mechanisms (Madani et al. 2011; Cascão and Nicol 2017). Non-cooperative game theory, which models players making independent decisions based on their interests, also falls short, as the absence of a clear Nash equilibrium underscores the difficulty of reaching mutually acceptable agreements when countries prioritize their own benefits over collective solutions (Lashitew & Kassa 2020b).

These challenges suggest that fairness and equitable distribution are significant stumbling blocks to resolution. To address these concerns, we turn to the concept of fair division, particularly the Steinhaus "cake-cutting" problem (Steinhaus 1948). By representing the Nile's water resources as a metaphorical cake to be divided among the riparian states, we can explore how different allocation algorithms, enhanced by AI, may lead to more equitable solutions. Our approach is innovative in that it integrates the classical Steinhaus fair division problem with modern AI algorithms, creating a unique framework for resolving resource allocation conflicts. This combination leverages Steinhaus's mathematical elegance and AI's computational power to offer practical, implementable solutions that address both strategic behaviors and normative concerns.

Building on the work of political philosopher John Rawls (1971) and his theory of justice, we establish a normative foundation for fairness in resource allocation. Rawls' principles, such as the "veil of ignorance" and the "difference principle," offer a framework for considering impartiality and the welfare of the least advantaged in the context of the Nile dispute. By integrating Rawlsian concepts with AI-driven fair division algorithms, we aim to identify potential solutions that align with principles of fairness while accounting for the strategic constraints and incentives faced by the riparian states.

The paper's methodology involves a sequential application of these concepts to the Nile water dispute. First, game theory is employed to analyze the strategic dynamics of the conflict and identify the limitations of purely game-theoretic approaches. Next, fair division algorithms, the Steinhaus cake-cutting problem, is introduced to explore equitable allocation solutions that account for the countries' diverse preferences and interests (Thomson, 2015).[1] Finally, some Rawlsian ideas are incorporated as a heuristic framework for evaluating the fairness of potential solutions and ensuring that the needs and concerns of all riparian states are considered.

This sequential methodology allows for a systematic examination of the Nile water dispute from multiple perspectives. By starting with game theory, the article establishes the strategic context of the conflict and highlights the challenges of reaching a stable agreement through purely strategic means. Introducing fair division algorithms provides a structured approach to exploring equitable allocation solutions beyond game theory's limitations. The incorporation of Rawlsian justice ensures that the proposed solutions are strategically viable and align with principles of fairness and equity.

---

[1] William Thomson uses axiomatic and game-theoretic analyses to understand complex problems. The relevance of our research on the Nile River dispute lies in the parallels between the concepts of fair division and the issues at hand. The Nile River dispute can be seen as a problem of fair division where multiple parties (countries) claim rights to a single, indivisible resource (the water of the Nile River). Thomson's work provides theoretical grounding for analyzing and potentially resolving such disputes.

Moreover, the sequential integration of these concepts creates a robust and coherent methodology for analyzing the Nile water dispute. Each step builds upon the insights gained from the previous step, allowing for a more nuanced understanding of the conflict and a more comprehensive exploration of potential solutions. The paper's methodology offers a novel and interdisciplinary approach to addressing complex resource allocation disputes by combining strategic analysis, algorithmic solutions, and normative considerations.

The article proceeds as follows: Section 2 introduces the Steinhaus cake-cutting problem and its relevance to the Nile dispute. Section 3 provides a literature review on the Nile conflict and the application of game theory and fair division concepts to water resource management. Section 4 presents a case study of the Nile dispute, focusing on the perspectives of Egypt, Sudan, and Ethiopia. Section 5 explores the limitations of cooperative and non-cooperative game theory in resolving the conflict. Section 6 demonstrates how cake-cutting algorithms, informed by Rawlsian principles, can offer insights into potential solutions. The article concludes by discussing the policy implications and the value of an interdisciplinary approach to resolving complex resource allocation disputes.

## 2. The Steinhaus "cake-cutting" Problem.

The Steinhaus cake-cutting problem, rooted in the work of mathematician Hugo Steinhaus, is a pivotal concept in fair division theory. It addresses the challenge of dividing a heterogeneous resource, metaphorically represented as a cake, among multiple parties with diverse preferences. The problem's complexity increases with the number of participants. In the context of the Nile River dispute, the cake represents the river's water resources, and the division of the cake signifies the allocation of these resources among the riparian countries Ethiopia, Egypt, and Sudan.

Rawls' theory of justice provides a philosophical foundation for key fairness principles in the cake-cutting problem. The "veil of ignorance," which involves concealing claimants' identities during allocation, aligns with the impartiality principle in fair division algorithms. This prevents self-interested behavior and promotes just resource divisions. Rawls' difference principle, tolerating inequality only if it benefits the least advantaged, is reflected in algorithms that maximize the minimum share allocated to any party.

Two essential notions of fairness emerge from Rawlsian ethics and fair division literature: proportionality and envy-freeness. Proportionality ensures each claimant receives at least a 1/n share when a resource is divided among n claimants, promoting equality in distribution. Envy-freeness, where no claimant prefers another's allocation over their own, guarantees that no party feels disadvantaged compared to others, upholding impartiality.

The cake-cutting problem can be represented as follows: Let the cake be the interval $[0, 1]$, and each agent i has a value density function $v_i(x)$ representing their subjective valuation of the cake at point x. The goal is to find an allocation $A = (A_1, ..., A_n)$, where $A_i$ is the portion allocated to agent i, so everyone believes they have the most valuable piece based on their preferences. The entire cake's value for an agent is represented by the integral of the value density function $v(x)$ over the interval $[0, 1]$, given by:

$$\int_0^1 v(x)dx$$

This value is assumed to be 1 for every agent, signifying each agent's valuation of the entire cake as 1.

The allocation $A=(A_1,...,A_n)$ showcases the cake's division among agents. Each agent *i* receives a segment $A_i$, which is mutually exclusive. The entire cake is the union of all slices, that is:

$$\bigcup_{i \in N} A_i = [0,1]$$

An allocation A is envy-free if, for all agents $i, j \in N$:

$$V_i(A_i) \geq V_i(A_j)$$

Every agent should value their segment at least as much as any other. No agent should envy another's piece. Classical cake-cutting literature posits that a protocol possesses a property like envy-freeness if every agent is assured not to feel envy by honestly participating in the protocol. This implies the protocol guarantees an envy-free division if all agents truthfully declare their value density functions.

The objective is to devise strategies and methods fostering collaboration and facilitating a fair, sustainable division of the Nile's water resources. To divide this resource equitably and metaphorically represent it as a cake, we utilize a game-theoretic algorithm considering both proportionality and envy-freeness. Every participant assigns a valuation function to the cake's sections, denoting their subjective value. We assume these functions to be non-negative and integrable over the interval.

3. A Literature Review

The Nile River basin is currently embroiled in significant water allocation disputes. The Nile, spanning 6,695 km, is one of Africa's most extensive and critical river basins, covering about 10% of the continent. Its pivotal role in the region's economy, environment, and social structures makes it a focal point in hydro-political discussions. Factors like population growth, economic expansion, and climate change amplify the Nile's importance. The escalating water demand, driven mainly by population growth, poses one of the most pressing challenges to finding equitable use of the water supplied from the Nile. The dispute involves Egypt and Ethiopia, centered on Ethiopia's construction of the Grand Ethiopian Renaissance Dam (GERD) on the Blue Nile (Yihdego, Rieu-Clarke, and Cascão 2017, Chapter 1). Related to this conflict is a long-standing disagreement between Egypt and Sudan over water use rights downstream (Lashitew & Kassa 2020a).

With the near completion of the dam, the dispute over GERD mainly revolves around how quickly its reservoir should be filled. While Ethiopia aims for a swift filling to optimize hydroelectricity production, downstream countries advocate for a more gradual approach, spanning 10-15 years, to mitigate adverse impacts. Any accelerated filling timeline could significantly affect Egypt, substantially reducing its irrigated farmlands and water supply. The negotiations have seen Egypt pushing for a guaranteed minimum annual water supply, a figure higher than what Ethiopia is willing to commit to.

After GERD's completion, Egypt's water inflow is expected to drop by at least 15%, necessitating sourcing 7.5-20 billion m3 of water from alternative means. The dam's implications have prompted strong reactions from Egyptian leaders, who regard the Nile as indispensable to Egypt's existence. While past leaders have issued stern warnings, the current leadership has adopted a more conciliatory approach, seeking harmony with Ethiopia and other Nile basin nations (Turhan, 2021: 75).

The nexus between resource scarcity and conflict has been a focal point among scholars (Zeitoun et al., 2020). Freshwater stands as a preeminent natural resource that nations increasingly compete for (Haftendorn, 2000). Factors such as population growth and agricultural expansion, especially in less economically developed nations like the Eastern Nile Basin, intensify this competition. Two perspectives dominate the literature surrounding escalating conflict over shared water resources.

The first perspective contends that heightened competition over freshwater resources invariably precipitates conflict among riparian nations. Esteemed scholars like Joyce Starr (1991), Thomas Homer-Dixon (1994), and John Waterbury have delved into the prospects of disputes and even wars stemming from burgeoning population growth and economic development. In his influential work (2002: 25-30), John Waterbury prognosticated a grim future due to Ethiopia's and Egypt's deep-seated national interests concerning the Nile. He delineated four crucial factors shaping the Nile Basin conflict.

The second perspective views water resources as conduits for future collaboration and shared security. Despite the evident conflict potential, over 300 treaties addressing shared waters between riparian nations showcase a trend toward cooperation. Arsano (2017) paints a vivid picture of Ethiopia's hydro-political intricacies in the Nile Basin, underscoring national and regional political complexities essential for understanding the ongoing GERD dispute. Research by Cascão and Nicol (2016), Salman (2019), and Yihdego and Rieu-Clarke (2016) provide insights into the Nile Basin's emerging cooperation norms, the progression to the Declaration of Principles, and fairness's exploration in international law concerning the Blue Nile and GERD. A conspicuous

void exists in the legal and institutional frameworks harmonizing upstream and downstream interests. The bilateral 1959 agreement between Egypt and Sudan notably omitted upstream nations from the negotiation and the accord.

This lack of harmonized water utilization and management has precipitated multifaceted challenges for the Eastern Nile's three riparian nations. Issues range from upstream erosion in Ethiopia and sediment accumulation in Sudan to excessive water evaporation in Egypt (Pemunta, 2021). Whittington, Wu, and Sadoff (2005) posited that regional collaboration is pivotal for watershed management, flood and drought mitigation, and sustainable power development (Lakew, 2020). Egypt's agricultural backbone faces dire threats without such cooperation, as elucidated by El-Nashar and Elyamany (2018).

From a normative perspective, which focuses on how things should ideally be, Egypt's refusal to reconsider the 1959 agreement, which predominantly benefits it as a downstream nation, appears unjust. This agreement does not account for the equitable rights of all riparian countries. Conversely, from a positive standpoint, considering how entities likely behave given certain conditions, Ethiopia's strategic position as the upstream nation grants it the leverage to regulate the Nile's flow. This control and its development objectives have driven Ethiopia to capitalize on its advantageous position.

Rawls' theory of justice provides some suggestive insights for evaluating the fairness of water allocation agreements from a normative perspective. Rawls argues that principles of justice should be chosen behind a 'veil of ignorance,' where parties are unaware of their position in society (Rawls, 1971). Concerning the Nile River dispute, this would mean considering the principles that riparian nations would agree to if they needed to know whether they were upstream or downstream countries. Rawls' principles of equal liberty and the 'difference principle,' which states that inequalities should be arranged to benefit the least advantaged, provide a foundation for assessing the fairness of water allocation agreements.

The literature on the Nile River dispute highlights the complex interplay of normative considerations, such as the equitable allocation of water resources, and the strategic realities the riparian nations face. Rawls' theory of justice provides a valuable framework for evaluating the fairness of water allocation agreements, emphasizing the importance of considering the needs and rights of all parties involved, particularly the least advantaged. However, the presence of power imbalances and the tendency towards non-cooperative behavior, as evidenced by Egypt's resistance to the GERD and Ethiopia's unilateral actions, underscore the challenges of achieving a just and stable resolution to the dispute (İlkbahar and Mercan's 2023). To further understand these complexities and explore potential strategies for resolution, we turn to the insights provided by game theory, particularly the concepts of cooperative and non-cooperative games (Mandani et al., 2011; Young, 1991).

## 4. The Nile River Dispute: An Exploration through Cooperative and Non-Cooperative Games

In game theory, we often speak of 'games' not in the traditional sense but as interactions where multiple players make strategic decisions that influence each other's outcomes. These games can be broadly categorized into cooperative and non-cooperative games. Cooperative games are scenarios where players can form binding agreements, and there's an expectation of negotiation in good faith to achieve a mutually beneficial outcome.

On the other hand, non-cooperative games involve players making decisions independently, often leading to suboptimal results, as each player seeks to maximize their benefit without considering the group's overall welfare. The Nile River dispute can be viewed as both a cooperative and non-cooperative game, depending on the context and the specific actions of the countries involved. This dual perspective allows us to explore a range of potential outcomes and strategies, from collaborative agreements to unilateral actions (Cascão 2009; Cascão and Nicol 2017, Chapter 5).

There is a long history of agreements and conflicts between countries regarding sharing Nile waters, often driven by Egypt's desire to protect its historical water rights (Badea, 2020; Matthews & Vivoda, 2023).

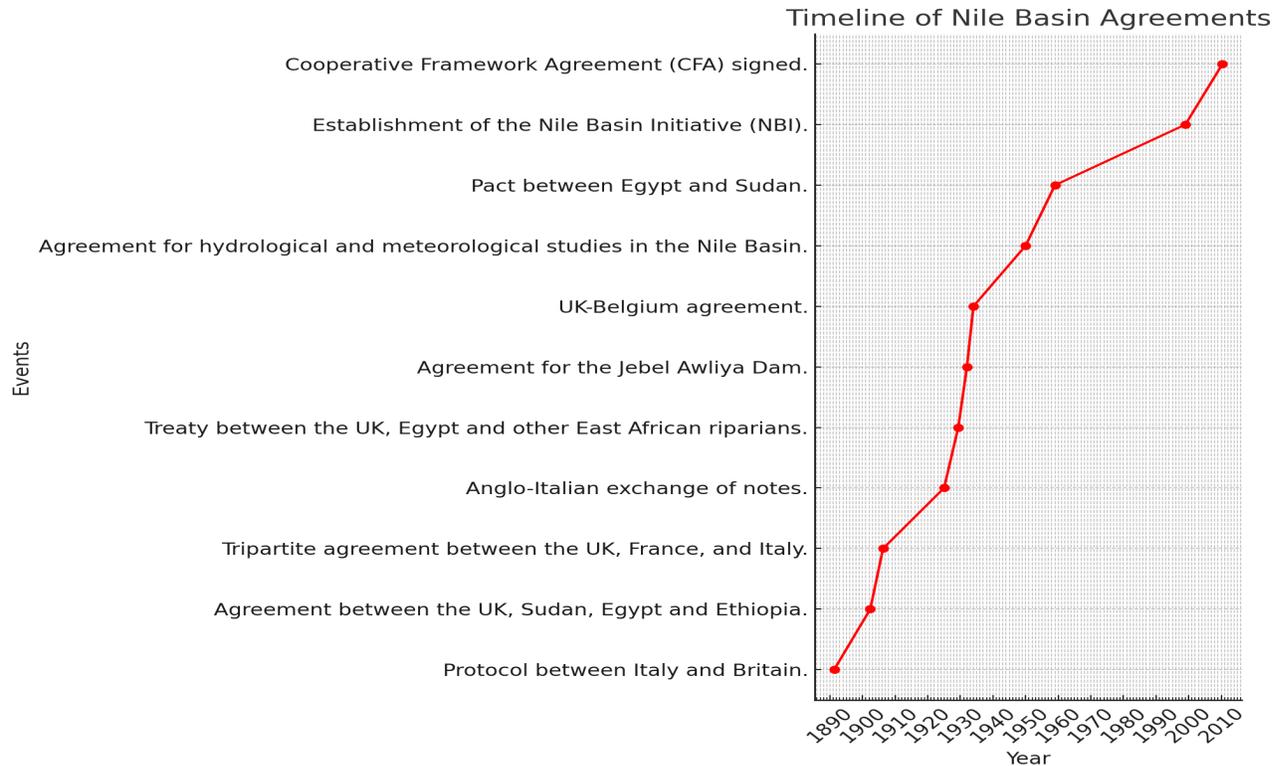

In 1929, Egypt and Britain (on behalf of Sudan and other British colonies) signed an agreement that allocated 48 billion cubic meters of Nile waters to Egypt and 4 billion to Sudan. This was followed in 1959 by an agreement between Egypt and Sudan that allocated 55.5 billion cubic meters to Egypt and 18.5 billion to Sudan, excluding other countries. Egypt has often threatened military action against upstream countries like Ethiopia and Sudan when they proposed projects affecting Nile waters (Tekuya, 2018). However, in the 1990s and 2000s, there were efforts for more cooperative management of the Nile, such as establishing the Nile Basin Initiative in 1999.

Despite periods of conflict, recent decades have seen increasing attempts at collaborative sharing of Nile waters between the riparian states (Dinar and Alemu, 2020; Jungudo, 2021). A cooperative agreement was reached in 2015, which called for further studies before Ethiopia could begin construction on the Nile; Ethiopia viewed it as a delaying tactic and proceeded with the dam project (Kimenyi and Mbaku 2015). Subsequently, negotiations regarding the dam have persisted for several years, involving various international mediators, such as the African Union, the United States, and the United Nations.

Negotiations have been marked by mistrust, disputes over technical matters, and broader geopolitical tensions (Tawfik, Rawia 2016). As the dam nears completion, Ethiopia has initiated the reservoir filling process despite objections from Egypt and Sudan. This escalation of tensions has generated urgent calls for diplomatic solutions to prevent the dispute from escalating into a full-blown conflict (Mbaku, 2020).

The Nile River dispute fundamentally boils down to a resource allocation issue: how can the Nile's waters be apportioned equitably among these countries? This is where the theoretical frameworks of cooperative and non-cooperative games come into play. In a cooperative game, the players can form binding agreements and negotiate in good faith to achieve a mutually beneficial outcome. Applied to the Nile River dispute, a cooperative approach would involve the three countries reaching a consensus on the dam's operation and the distribution of the Nile's waters (Wiebe, 2001). However, the ongoing negotiations, marked by mistrust and disagreements, suggest that the conditions for a cooperative game still need to be fully met. A non-cooperative game involves players making decisions independently, often leading to suboptimal outcomes. The current state of the Nile River dispute, with Ethiopia unilaterally proceeding with the dam's construction despite objections from Egypt and Sudan, resembles a non-cooperative game.

The non-cooperative aspect persists despite efforts by outside parties to mediate. The African Union has facilitated negotiations between Egypt, Sudan, and Ethiopia since 2007, leading to a 2010 Cooperative Framework Agreement for equitable use of Nile waters. However, AU's influence remains limited, and binding agreements have not been reached. The World Bank created initiatives in the late 1990s and 2000s to support cooperation and conflict resolution between Nile riparians. It has provided funds, expertise, and facilitation of working groups and ministerial dialogues. However, the lack of a formal dispute resolution mechanism and comprehensive basin-wide management plan has limited progress.

The European Union has assisted with developing the Cooperative Framework Agreement since the 2000s through financial and technical support. The US attempted to broker agreements in 2016-2020 but was unsuccessful due to resistance from Ethiopia. US provision of financial aid to Ethiopia and Sudan had yet to do much to resolve the dispute. At Sisi's request, the Trump administration got involved through its Secretary of Finance, but this interference only hardened Ethiopia's position. Third-party facilitation has yielded some cooperation, but results are limited. Fundamental tensions remain unresolved (Tuhran, 2021: 74).

## 4.1 Non-cooperation as a Dominant Strategy

This is where concepts from game theory can provide valuable insights. The Nile River dispute is a prime example of a complex resource allocation problem. Each country involved has its own needs, interests, and perceptions of what constitutes a fair share of the Nile's waters. diverse perspectives while ensuring the sustainable and equitable use of the river. The non-cooperative logic that prevails is illustrated in the game model below. The payoff matrix represents the outcomes for each country under different strategy combinations.

**Game matrix 1.**

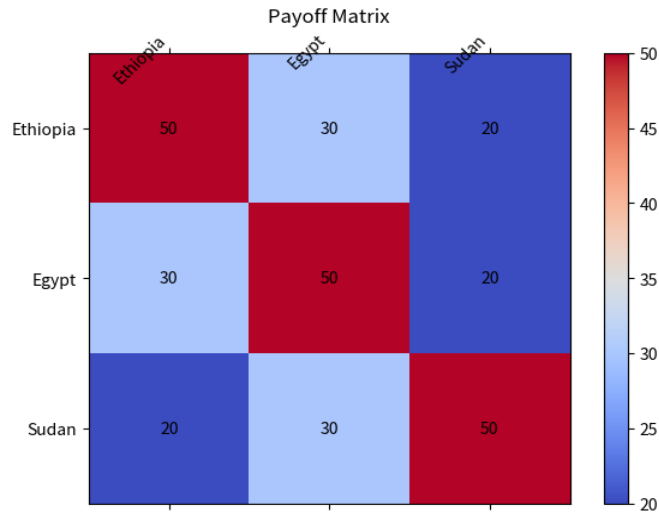

The matrix reveals Ethiopia's dominant strategy is to fill the GERD quickly (E1) to maximize its benefits from hydropower generation and water resource control. Conversely, Egypt's dominant strategy is to resist the dam (A2) to protect its historical water rights and maintain its current allocation of Nile waters. The intersection of these dominant strategies leads to a Nash equilibrium at (E1, A2), where no country can unilaterally improve its outcome by changing its strategy, given the strategy of the other country. This non-cooperative Nash equilibrium does not lead to an optimal outcome for the riparian countries. The equilibrium outcome (30, 20, 50) reflects the countries' pursuit of their interests.

The challenges in reaching a cooperative solution arise from the countries' misaligned objectives and the potential for unilateral actions to harm others. If Ethiopia proceeds with the rapid filling of the GERD, it could significantly reduce water flow to Egypt and Sudan, prompting retaliatory measures. On the other hand, strong opposition from Egypt and Sudan could result in a stalemate, with all countries entrenched in their positions and unable to reach a mutually beneficial agreement. In contrast, a cooperative outcome would involve the countries negotiating a water allocation plan that maximizes their combined welfare.

The second model represents a cooperative solution in which Ethiopia slowly fills the GERD, and Egypt and Sudan accept the dam. This cooperative outcome leads to a more equitable distribution of the Nile's waters and a higher total payoff for the riparian countries.

**Game matrix 2.**

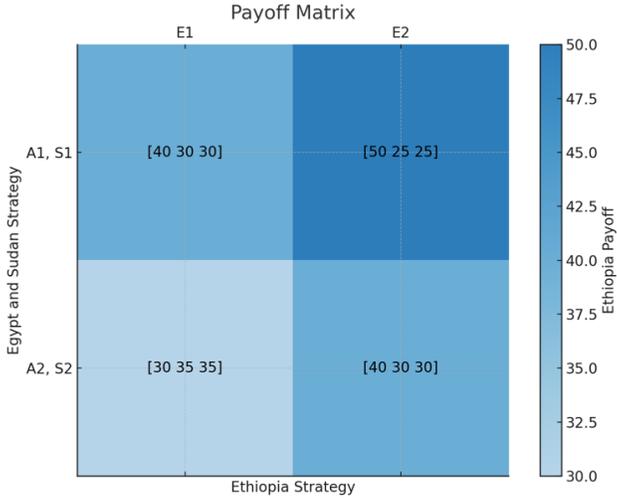

However, the divergent strategic interests among the countries make achieving and sustaining a cooperative solution difficult. The absence of an effective enforcement mechanism and the temptation for governments to prioritize their interests over collective welfare render a cooperative Nash equilibrium unlikely (Madani et al., 2011; Brânzei and Miltersen, 2013).

The non-cooperative nature of the dispute was evident in the failed negotiations between Egypt, Sudan, and Ethiopia in August 2023. Despite Egypt's insistence on a legally binding agreement to protect its water rights, Ethiopia maintained that the GERD is a matter of national sovereignty and non-binding recommendations would suffice (Hendawi, 2023). The core disagreements, including competing claims about the dam's potential impact on water flows, continue to hinder progress toward a cooperative solution.

The game theory models reveal that the Nile River dispute is characterized by dominant non-cooperative strategies, where countries pursue their dominant strategy at the expense of collective welfare. Transitioning from this non-cooperative equilibrium to a cooperative solution requires addressing fair and equitable water allocation and introducing mechanisms that promote cooperation and discourage unilateral actions. In the following section, we will explore how the principles of fair division, such as the Steinhaus cake-cutting problem and the "I cut, you choose" algorithm, can be applied to the Nile River dispute to identify potential pathways for achieving a cooperative and equitable solution.

## 5. Applying the "Cake-Cutting" Problem to the Nile Dispute

To tackle the challenge of fairly distributing the Nile's resources, we must explore alternative approaches beyond the limitations of non-cooperative game theory. One such approach is the Steinhaus "cake-cutting" problem, a mathematical framework for solving fair division problems. The principles of the cake-cutting problem offer new perspectives on allocating the river's waters in a way that satisfies fairness criteria such as proportionality and envy-freeness.

We consider three countries: Egypt, Sudan, and Ethiopia, each with its preferences for how the Nile's resources should be divided. This allocation can be based on the Selfridge Conway discrete procedure, which guarantees envy-freeness. In this procedure, Player 1, the first step divider receives the trimmed part (Piece 0) and Piece 2. Player 2, the trimmer in the second step, receives Piece 0, and Player 3, the chooser in the third step, selects Piece 1. This allocation ensures that each player believes they have received the most valuable piece(s) according to their valuation. This is the essence of envy-freeness: no player envies another player's piece of cake.

The table below shows the outcome of this fair division algorithm, highlighting that each player believes they have received the most valuable piece(s) according to their valuation, fulfilling the condition of envy-freeness. The specific pieces were allocated to ensure satisfaction for each player based on their valuations. Such an allocation method can be adapted to real-world resource divisions, such as those in international river disputes, ensuring that all parties perceive the distribution as fair.

**Table 1.**

| Player | Allocated Piece(s) |
|---|---|
| Player 1 | Piece 0 (trimmed part) and Piece 2 |
| Player 2 | Piece 0 |
| Player 3 | Piece 1 |

This table illustrates the outcome of the fair division algorithm. Player 1, who acts as the divider, receives the trimmed part and Piece 2. Player 2, the trimmer, receives Piece 0, while Player 3, the chooser, gets Piece 1. This distribution ensures that each player perceives their allocation as the most valuable, adhering to the principle of envy-freeness.

The value density functions for Ethiopia, Egypt, and Sudan illustrate how each country values different parts of the Nile's resources. The x-axis represents the resource (Nile River's water), and the y-axis represents the value density, indicating each country's valuation of different resource portions. Ethiopia's value density function, represented by a sine wave, shows a higher valuation for the middle segments of the resource, reflecting its interest in hydroelectric projects or other specific uses. Egypt's value density function, represented by a cosine wave, indicates a stable valuation across the resource but a higher valuation for the initial segments, reflecting its dependence on upstream flow for agriculture and drinking water. Sudan's value density function, a sine wave with a higher frequency, indicates fluctuating valuations, reflecting its complex position balancing electricity supply, flood management, and irrigation needs.

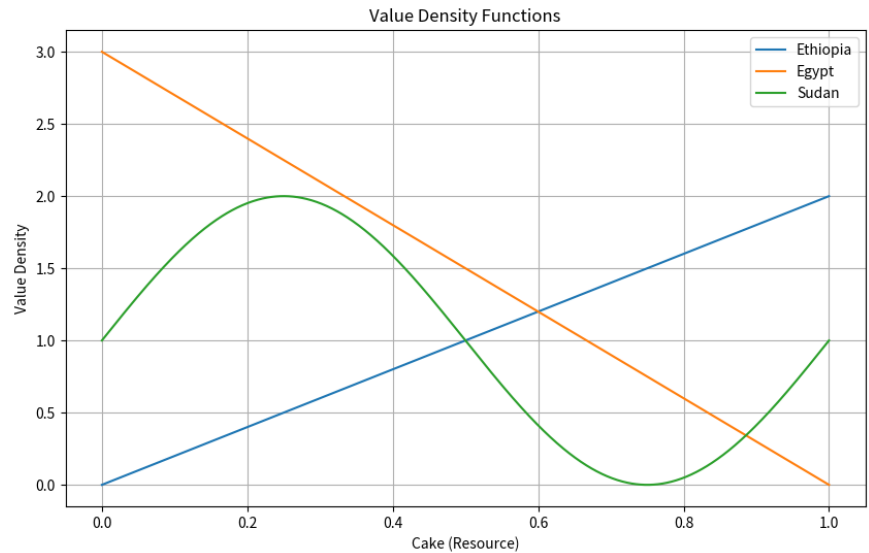

These valuation differences complicate agreements, but a fair division algorithm like cake-cutting can help address these differences by aiming for an allocation that all parties perceive as fair. Understanding these value density functions is crucial for implementing the cake-cutting algorithm, ensuring that each country's segment aligns with their highest valuation, supporting fairness and envy-freeness. This analysis supports the broader goal of equitable and sustainable water management in the region.

The Adjusted Winner algorithm ensures an envy-free allocation where no participant prefers another's share over their own. The steps involved are: each country assigns specific points to different parts of the resource, reflecting their respective valuations. The total points assigned by each country are equal, ensuring fairness. The algorithm initially allocates each part to the country assigning the most points. Further adjustments transfer parts of the resource from the country with more points to balance the allocation. In the Nile Basin context, resources are divided into ten intervals and distributed among the three countries based on their value density functions. This allocation is sorted in descending order, with larger intervals allocated first. The sum of intervals represents each country's total share of the Nile's resources.

The value density functions for each country reveal their preferences and priorities, which are crucial for implementing fair division algorithms like the Adjusted Winner algorithm. For example, Ethiopia prioritizes hydroelectric power, reflecting a higher valuation for the later parts of the river. Egypt focuses on maintaining a reliable water flow for agriculture and drinking water, reflected in a higher valuation for the earlier parts. Sudan displays fluctuating valuations, indicating a complex position balancing electricity supply, flood management, and irrigation needs.

Applying the Adjusted Winner algorithm in the Nile Basin context holds several policy implications. Firstly, it promotes fairness and equity in resource allocation among the countries involved. By considering individual valuations and striving for an envy-free allocation, policymakers can address concerns about inequality and foster cooperation among the countries in the region. Secondly, the algorithm provides a structured framework for negotiations and decision-making. Policymakers can use the algorithm for constructive dialogue and agreements, considering each country's varying needs and priorities. Engaging in inclusive and transparent discussions, policymakers can work towards sustainable resource management that balances economic, social, and environmental considerations. Moreover, the algorithm's adaptive nature encourages policymakers to continually assess and adjust allocations based on changing circumstances, facilitating flexible and adaptive decision-making.

While the Adjusted Winner algorithm effectively ensures envy-freeness, it is crucial to recognize that other desirable properties like efficiency and equitability may not be guaranteed. Envy-freeness is a scenario in which no participant would prefer to switch their allocation with another participant's. In complex contexts like the Nile Basin, achieving complete satisfaction and envy-freeness may still be challenging due to the nations' diverse interests and strategic behaviors.

**5.2 "I Cut; You Choose."**
The "I Cut; You Choose" algorithm is a simple method derived from the Steinhaus "cake-cutting" problem. It involves one person dividing the 'cake' into what they perceive as equal parts, and the others then choosing their preferred part. This method encourages the 'cutter' to divide the 'cake' as equally as possible, as they do not know which part they will receive (Kyropoulou, Ortega, and Segal-Halevi, 2022).

Let's denote the three parties involved as follows:

E: Ethiopia

Eg: Egypt

S: Sudan

Assume that the resource to be divided (water from the Nile River) is normalized and represented as the interval [0,1].

As the cutter, Ethiopia splits the interval into three subintervals, which it perceives as equal in value. Let's denote these subintervals as $I_E$, $I_{Eg}$, and $I_S$.

Egypt and Sudan, as the choosers, choose their preferred intervals based on their valuations. Let's denote their choices as $C_{Eg}$ and $C_S$, respectively.

If Egypt and Sudan choose different intervals (i.e., $C_{Eg} \neq C_S$), then each country gets the interval it chose, and Ethiopia gets the remaining interval.

If Egypt and Sudan choose the same interval (i.e., $C_{Eg}=C_S$), a conflict resolution mechanism is activated. This could involve another round of cutting and choosing or an external mediator making the final decision.

When Ethiopia is the cutter, the payoffs are: Ethiopia: 0.333, Egypt: 0.333, Sudan: 0.333

When Egypt is the cutter, the payoffs are: Ethiopia: 0.333, Egypt: 0.333, Sudan: 0.333

When Sudan is the cutter, the payoffs are: Ethiopia: 0.333, Egypt: 0.333, Sudan: 0.333

The bar plots graphically display these payoffs. In each case, the payoff is divided equally among the three countries.

In each case, the "I Cut; You Choose" algorithm leads to an equal resource division, irrespective of who the cutter is. This suggests that the algorithm can achieve a fair and equitable allocation of the Nile's waters among the three countries. However, this assumes that the cutter can divide the resource into equal parts based on their valuation and that the choosers can select their most preferred part. These assumptions may not hold due to the complex and strategic nature of the Nile River dispute. It's also important to note that this model doesn't consider the strategic behavior of the cutter, who may not divide the cake equally if they anticipate the choices of the other countries.

**Histogram 1.**

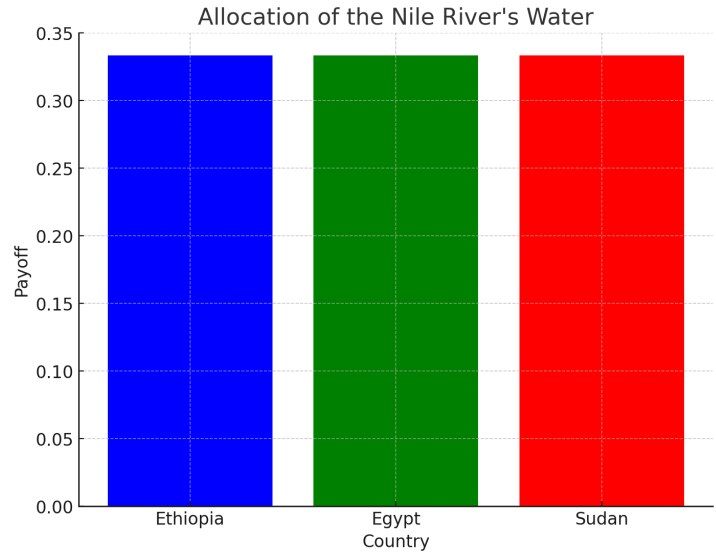

As you can see, under the "I Cut; You Choose" algorithm, with Ethiopia as the cutter, each country receives an equal share of the Nile River's water. However, it is essential to note that this is under the assumption that Ethiopia, as the cutter, splits the resource in a way that it perceives as fair. This may not necessarily be the case, as Ethiopia might have different priorities or strategic considerations influencing its decision.

Furthermore, as the choosers, Egypt and Sudan might not necessarily agree with Ethiopia's perception of a fair split. They might have different valuations for different resource allocations, which could lead to disagreements or conflicts. In a strategic setting, the "I Cut; You Choose" algorithm offers a potential way to achieve a fair and mutually acceptable allocation of a disputed resource. Thus, its success depends on the countries' willingness and ability to negotiate in good faith and respect the process's outcome.

**Histogram 2.**

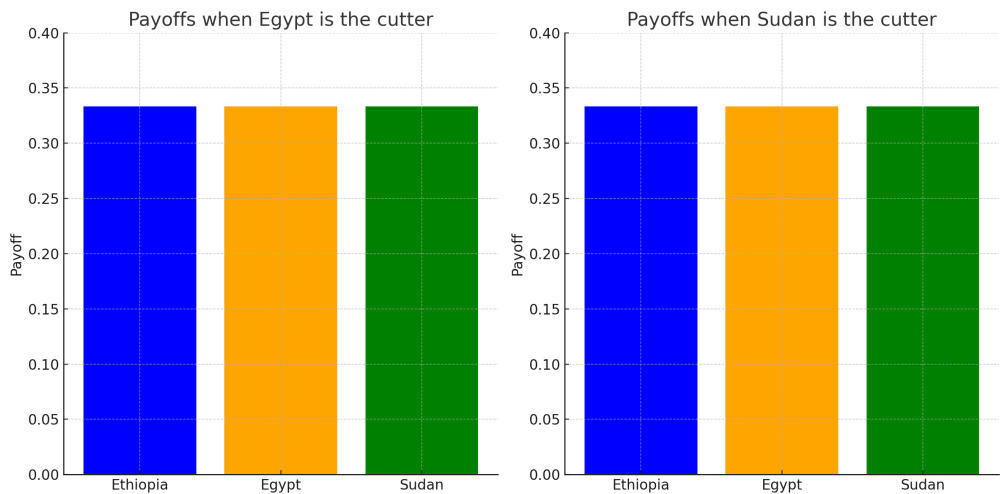

Histogram 2 represents the payoffs for each country when they act as the cutter in the "I Cut; You Choose" algorithm. This visualization helps to understand how the distribution of resources changes depending on which country is dividing the resources. As we can see, each country obtains the highest payoff when they are the ones dividing the resources, reflecting that they divide the resources in a way they perceive as most fair. However, the payoff for the other countries varies depending on the cutter, indicating that their perception of fairness may differ.

While the "I Cut; You Choose" algorithm provides a practical mechanism for fair division, its principles can be aligned with Rawls' theory of justice. Rawls' veil of ignorance posits that fair decisions are made when individuals do not know their societal position. Similarly, the "I Cut; You Choose" method ensures that the cutter is incentivized to divide the resource equitably, not knowing which part they will receive. This operationalization of Rawlsian fairness promotes equitable outcomes by reducing the potential for bias and self-interest, embodying the principles of fairness and impartiality essential for just resource allocation.

**5.3 They'll Be Dam(n) to Cooperate**

The "I Cut; You Choose" algorithm, when each country gets a chance to be the cutter, provides an exciting and potentially helpful framework for exploring solutions to the Nile River dispute. Incorporating each country's preferences and perceptions into the resource allocation process can help promote fairness, transparency, and mutual understanding among the nations. As with any theoretical model, its applicability and effectiveness in the real world would depend on a range of factors, including the willingness of the countries to engage in good faith.

From matrix 3, we can see that the payoffs vary depending on who is the cutter. This reflects that each country has different preferences and perceptions of what constitutes a fair share of the Nile's waters.

**Game Matrix 3.**

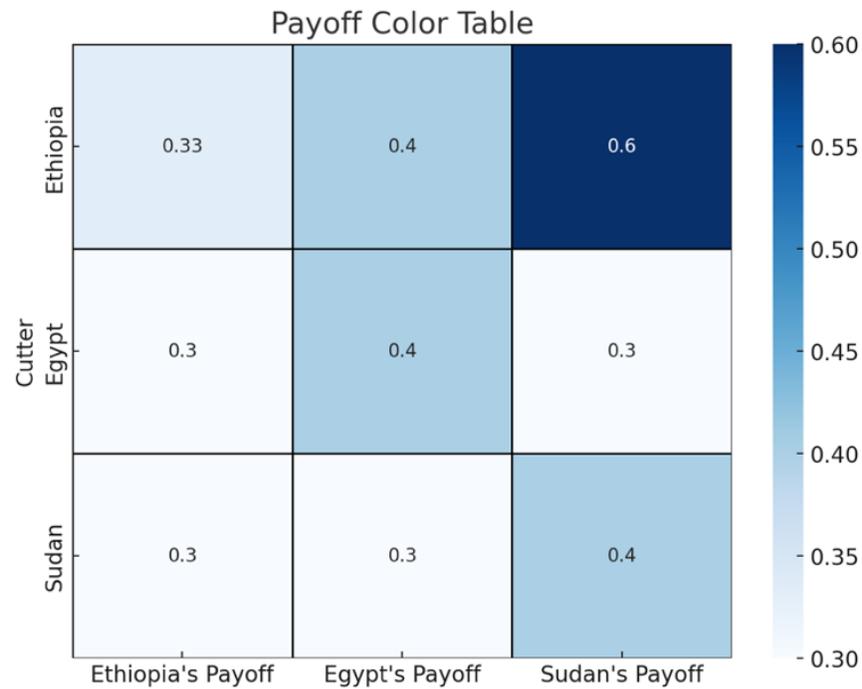

When each country is given the chance to be the cutter, it divides the resources in a way it perceives as fair, but this may need to align with the other countries' perceptions. This is the essence of the "I Cut; You Choose" algorithm: it strives to achieve a fair allocation from the cutter's perspective. However, the allocation's actual fairness may depend on the preferences and perceptions of other countries.

Bar Chart 1.

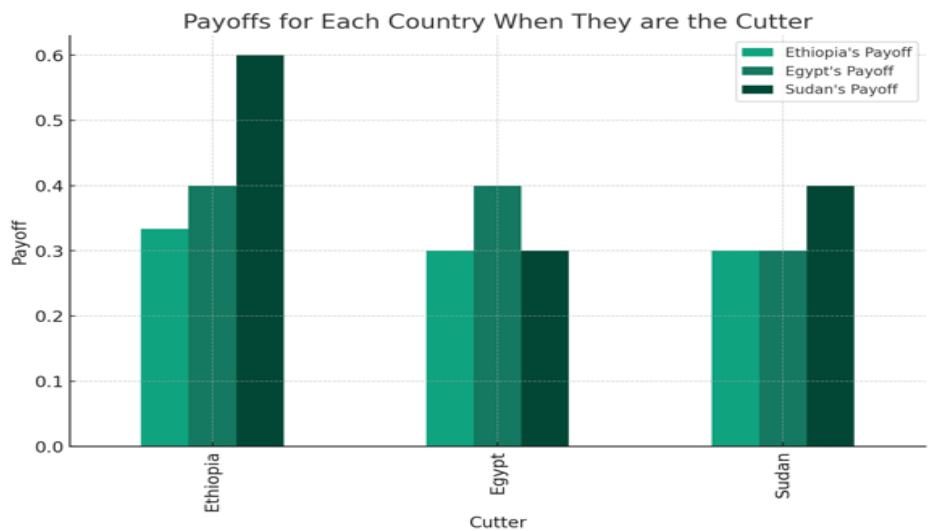

Interestingly, the "I Cut; You Choose" algorithm brings an element of fairness into the distribution of resources, even in a context where cooperative game theory initially failed. Allowing each country the opportunity to act as the cutter encourages consideration of a fair distribution from the perspective of all parties involved. This cooperative element is embedded within a non-cooperative setting, requiring countries to contemplate how the others might act in the subsequent choosing phase. Although this algorithm doesn't directly solve the conflict or force cooperation, it creates a structure that might nudge actors toward more cooperative strategies.

The "I Cut; You Choose" algorithm shares similarities with John Rawls's influential theory of "justice as fairness," rooted in principles of impartiality, equality, rational self-interest, procedural justice, and maximizing benefits for the least advantaged. Specifically, the algorithm's impartial approach of the cutter not knowing which portion they will receive mirrors Rawls's "veil of ignorance" thought experiment for unbiased decision-making.

Furthermore, based on the cutter's perspective, the algorithm promotes fairness by seeking to divide the resource into equally valued portions. Rawls emphasized equality as the foundation for justice but allowed for inequalities if they benefited the least advantaged. The cutter, driven by rational self-interest, aims to divide portions fairly to avoid the least preferred portion, paralleling Rawls's point that self-interest under impartial conditions leads to just outcomes.

The procedural fairness of the algorithm's rules matters more than the final allocation, aligning with Rawls's focus on just processes over outcomes. Finally, the cutter seeks to maximize the minimum portion's value via a maximin rule, comparable to Rawls's argument that justice requires maximizing benefits for the least-advantaged individual. In summary, the parallels between the algorithm and Rawls's work highlight how these fundamental principles can foster mutually acceptable and fair solutions amidst conflicting interests (Rawls, 1971).

However, Brams, Jones, and Klamler (2011) strongly suggest no perfect solution exists. In their essay, "N-Person Cake-Cutting: There May Be No Perfect Division," they examine the theoretical problem of fairly dividing a heterogeneous, divisible resource (represented as a cake) among n players. The paper focuses on three properties of fair division: efficiency, envy-freeness, and equitability. In their definition, a "perfect" division satisfies all three properties. They provide an example of three players who find it impossible to find a perfect division, no matter how many cuts are allowed. An efficient and equitable division can be found with two cuts, but it is not envy-free. With three cuts, an equitable and envy-free division exists, but it is inefficient. The 4-cut allocation is efficient and equitable but not envy-free.

Thus, only two properties are achievable at once. The impossibility is proven for any number of players, not just 3. No perfect division exists for the given valuation functions, regardless of players and cuts allowed. When perfection can't be attained, the allocation that maximizes total player value could be chosen as the next-best solution. However, neither envy-freeness nor equitability dominates the maximization of value.[2]

Nonetheless, introducing a process where each country must consider others' responses requires strategic thinking beyond maximizing one's gain. Each country, acting as the cutter, is incentivized to divide the resource in a way perceived as fair, knowing that the others will then choose their portions. This dynamic can encourage compromise and foster a deeper understanding of the other countries' perspectives, thereby fostering a shift towards a more cooperative mindset. Moreover, the iterative nature of the algorithm, where each country has the chance to be the cutter, can help to balance power asymmetries.

---

[2] In recent work, they argue, "by having the players submit their value functions over the cake to a referee rather than move knives according to these functions the referee can ensure that the division is not only envy-free but also maximin. In addition, the referee can use the value functions to find a maximally equitable division whereby the players receive equally valued maximal shares. Still, this allocation may not be envy-free" (Brams, Jones, and Klamler (2022).

In the context of the Nile River dispute, where power dynamics play a significant role, this aspect of the algorithm can be particularly relevant. By placing each country in the position of the cutter, the algorithm ensures that each country's preferences and interests are considered in the resource allocation process. Thus, "I Cut; You Choose" algorithm is a potential tool to transition from a non-cooperative game, which has led to a stalemate, to a more cooperative approach.

## 6. Policy Implications and Conclusion

Our exploration of the Nile River dispute through various theoretical lenses makes it clear that innovative approaches such as the cake-cutting algorithm are theoretically robust and hold substantial practical implications. Comparing this approach with traditional conflict resolution models — bargaining models, utility maximization models, mediation, and arbitration — underscores its unique potential to transform complex disputes.

Unlike traditional bargaining models that may favor stronger parties or utility maximization models that focus on overall efficiency at the potential cost of individual fairness, the cake-cutting algorithm ensures that each party perceives its share as fair. This directly addresses the core challenges of envy-freeness and proportionality, which are crucial for long-term conflict resolution. Additionally, by engaging each party actively in the resolution process, the algorithm minimizes strategic manipulations and promotes more efficient negotiations.

The adaptability of the cake-cutting algorithm to multiple stakeholders with varied interests offers a significant advantage in managing complex disputes like those over the Nile waters. While the implementation requires careful setup to ensure accurate representation of parties' valuations, the transparency and simplicity of the process facilitate broader acceptance and adherence to the outcomes. With its foundation in fairness and clarity, the cake-cutting algorithm is likely to be more readily accepted by all parties involved than other models. However, for such theoretical models to be practically effective, supportive policy frameworks and legal provisions must ensure enforceability and compliance.

These insights into the cake-cutting algorithm's comparative advantages should inform the development of policy recommendations for the Nile dispute and similar international conflicts. Policymakers and negotiators are encouraged to consider these innovative methodologies, which promise theoretical fairness and offer practical pathways to achieving lasting resolutions.

Integrating game theory, fair division algorithms, and insights from political philosophy into conflict resolution practices offers a novel interdisciplinary approach. This approach should be further explored and developed in policy frameworks, enriching the dialogue around conflict resolution and opening avenues for equitable and sustainable solutions in international disputes (Young, 1991; Bueno de Mesquita, 2017).